\documentclass[conference]{IEEEtran}
\IEEEoverridecommandlockouts %
\usepackage{cite}
\usepackage{graphicx}%
\usepackage{url}
\usepackage{cite}
\usepackage{graphicx}
\usepackage{amsmath}
\usepackage{amssymb}
\usepackage{subfig}
\usepackage{lipsum, color}
\usepackage{mathtools, cuted}
\usepackage{textcase}

\newcommand{\ba}{\begin{array}}
\newcommand{\ea}{\end{array}}
\newcommand{\be}{\begin{displaymath}}
\newcommand{\ee}{\end{displaymath}}
\newcommand{\ben}{\begin{equation}}
\newcommand{\een}{\end{equation}}
\newcommand{\bena}{\begin{eqnarray}}
\newcommand{\eena}{\end{eqnarray}}
\newcommand{\beqa}{\begin{eqnarray*}}
\newcommand{\enqa}{\end{eqnarray*}}

\newcommand{\bc}{\begin{center}}
\newcommand{\ec}{\end{center}}
\newcommand{\bi}{\begin{itemize}}
\newcommand{\ei}{\end{itemize}}
\newcommand{\benu}{\begin{enumerate}}
\newcommand{\eenu}{\end{enumerate}}
\newcommand{\bdes}{\begin{description}}
\newcommand{\edes}{\end{description}}
\newcommand{\bt}{\begin{tabular}}
\newcommand{\et}{\end{tabular}}

\newcommand{\circlambda}{\mbox{$\Lambda$
             \kern-.85em\raise1.5ex
             \hbox{$\scriptstyle{\circ}$}}\,}

\begin{document}

\title{\huge Quantum Feature Extraction for THz Multi-Layer Imaging}
\author{
\IEEEauthorblockN{
T.~Koike-Akino\IEEEauthorrefmark{1}, 
P.~Wang\IEEEauthorrefmark{1}, 
G.~Yamashita\IEEEauthorrefmark{2}, 
W.~Tsujita\IEEEauthorrefmark{2},  
and M.~Nakajima\IEEEauthorrefmark{3}}
\IEEEauthorblockA{
\IEEEauthorrefmark{1}Mitsubishi Electric Research Laboratories (MERL), Cambridge, MA 02139, USA.
\\
\IEEEauthorrefmark{2}Mitsubishi Electric Corporation Advanced Technology R$\&$D Center, Amagasaki City, 661-8661, Japan.
\\
\IEEEauthorrefmark{3}Institute of Laser Engineering, Osaka University, Osaka 565$-$0871, Japan.}
}

\maketitle

\begin{abstract}
	A learning-based THz multi-layer imaging has been recently used for contactless three-dimensional (3D) positioning and encoding. 
	We show a proof-of-concept demonstration of an emerging quantum machine learning (QML) framework to deal with depth variation, shadow effect, and double-sided content recognition, through an experimental validation. 
\end{abstract}
	
\IEEEpeerreviewmaketitle
	
\section{Introduction}
	
\begin{figure}[t]
		\centering
		\subfloat[][THz-TDS multi-layer imaging]{
			\includegraphics[width=0.42\linewidth]{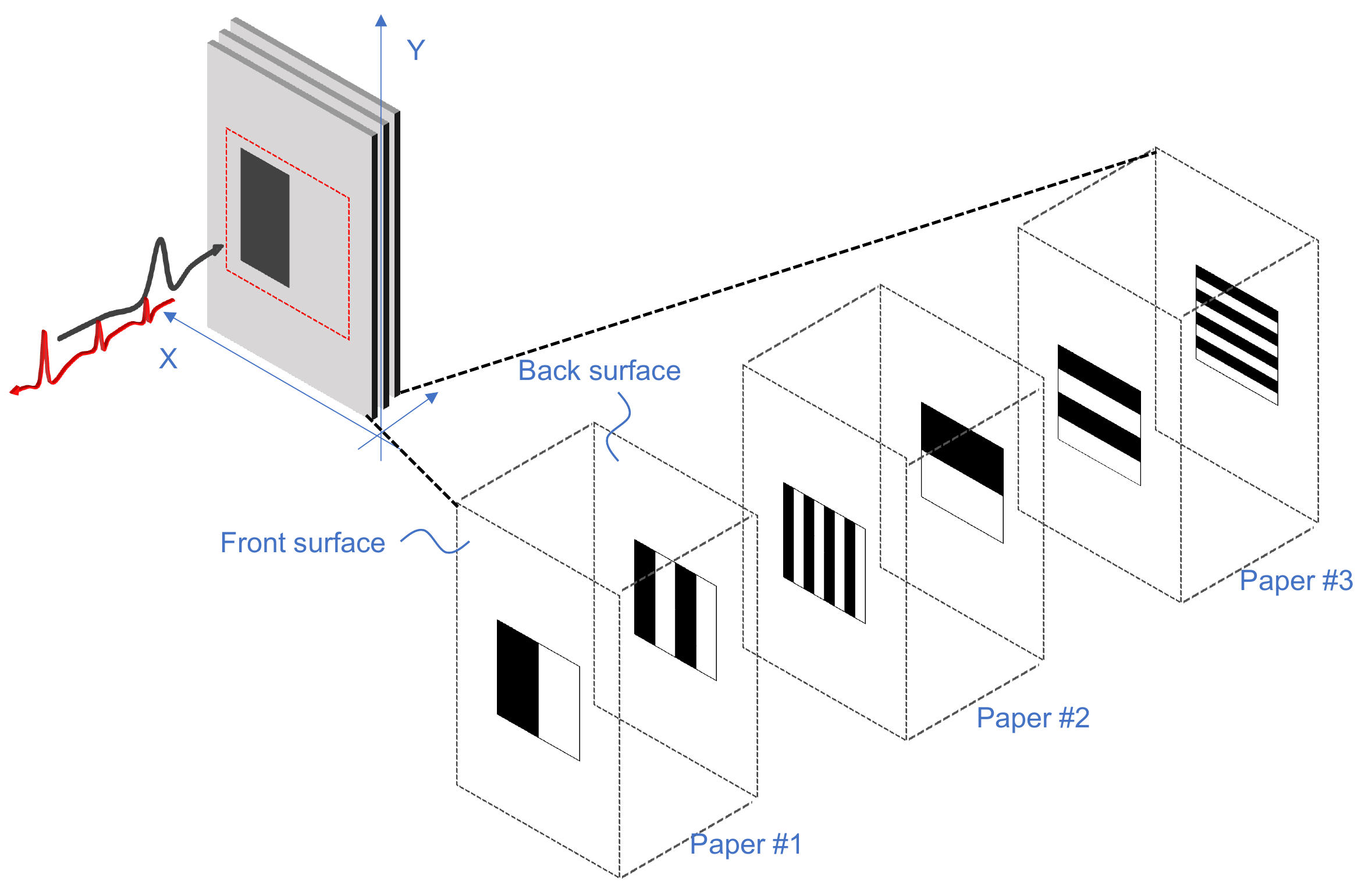}
		}
		\hfil
		\subfloat[][Shaddow effect]{
			\includegraphics[width=0.52\linewidth]{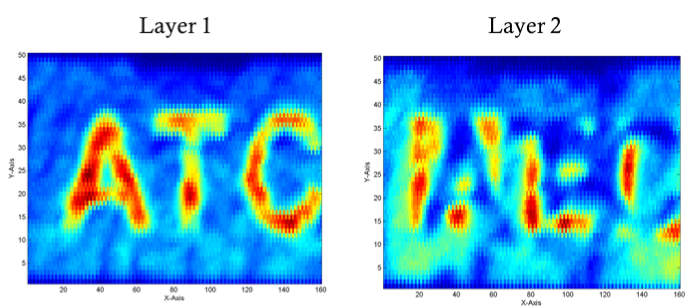} 
		}
	\caption{(a) THz-TDS 3D positioning with multi-layer 2D encoders;  
	(b) the shadow of three letters on the 1st layer is seen on the 2nd layer. }
	\label{fig:scan}		 
	\end{figure}

	The use of terahertz (THz) wave has drawn much attention for various industrial applications due to contactless sensing, usability under adversarial conditions (e.g., fire and smoke), and robustness to dust and dirt~\cite{WangSadeqi18, SadamotoTsujita18, YamashitaTsujita20, FuWang18, WangKoike20, WangKoike19, WangKoike21}. 
	We consider THz positioning in a \emph{raster} scanning mode~\cite{WangSadeqi18, SadamotoTsujita18, YamashitaTsujita20} to recognize three-dimensional (3D) internal layers as shown in Fig.~\ref{fig:scan}(a). 
	
	Compared with 1D~\cite{WangSadeqi18, YamashitaTsujita20, SadamotoTsujita18} and 2D encoders~\cite{FuWang18, WangKoike19, WangKoike20}, the multi-layer 3D THz encoders~\cite{WangKoike21} can increase capacity by exploiting THz penetration capability through non-conducting materials. 
	Nevertheless, we need to deal with practical challenges: 1) depth variation due to the platform vibration; 2) shadow effect caused by non-uniform penetrating illumination from front layers to deep layers; and 3) content recognition in the back surface of each layer. 
	Recently, deep learning has shown promising results to tackle those issues~\cite{WangKoike19, WangKoike21}.

	In this paper, we further extend it by introducing an emerging quantum machine learning (QML) framework~\cite{cerezo2021cost}, which offers exponentially rich state representation for feature extraction.
	To the best of our knowledge, this is the very first proof-of-concept paper of QML applied to THz sensing.

\section{QML-Assisted THz Multi-Layer Imaging}

\subsection{Quantum Machine Learning (QML)}
	An emerging QML framework has been recently applied to variout applications including sensing and communications~\cite{koike2022quantum,koike2022autoqml,liu2022learning}, envisioning future era of quantum supremacy.
	Quantum computers have the potential to realize computationally efficient signal processing compared to traditional digital computers by exploiting quantum mechanism, e.g., superposition and entanglement, in terms of not only execution time but also energy consumption.
	In the past few years, several vendors have successfully manufactured commercial quantum processing units (QPUs).
	For instance, IBM released $127$-qubit QPUs in 2021, and plans to produce $1121$-qubit QPUs by 2023.
	Recently, hybrid quantum-classical algorithms based on the variational principle were introduced as a state-of-the-art QML method to deal with quantum noise.

\subsection{Proposed QML Method}
	
\begin{figure}[t]
		\centering
		\includegraphics[width=0.9\linewidth]{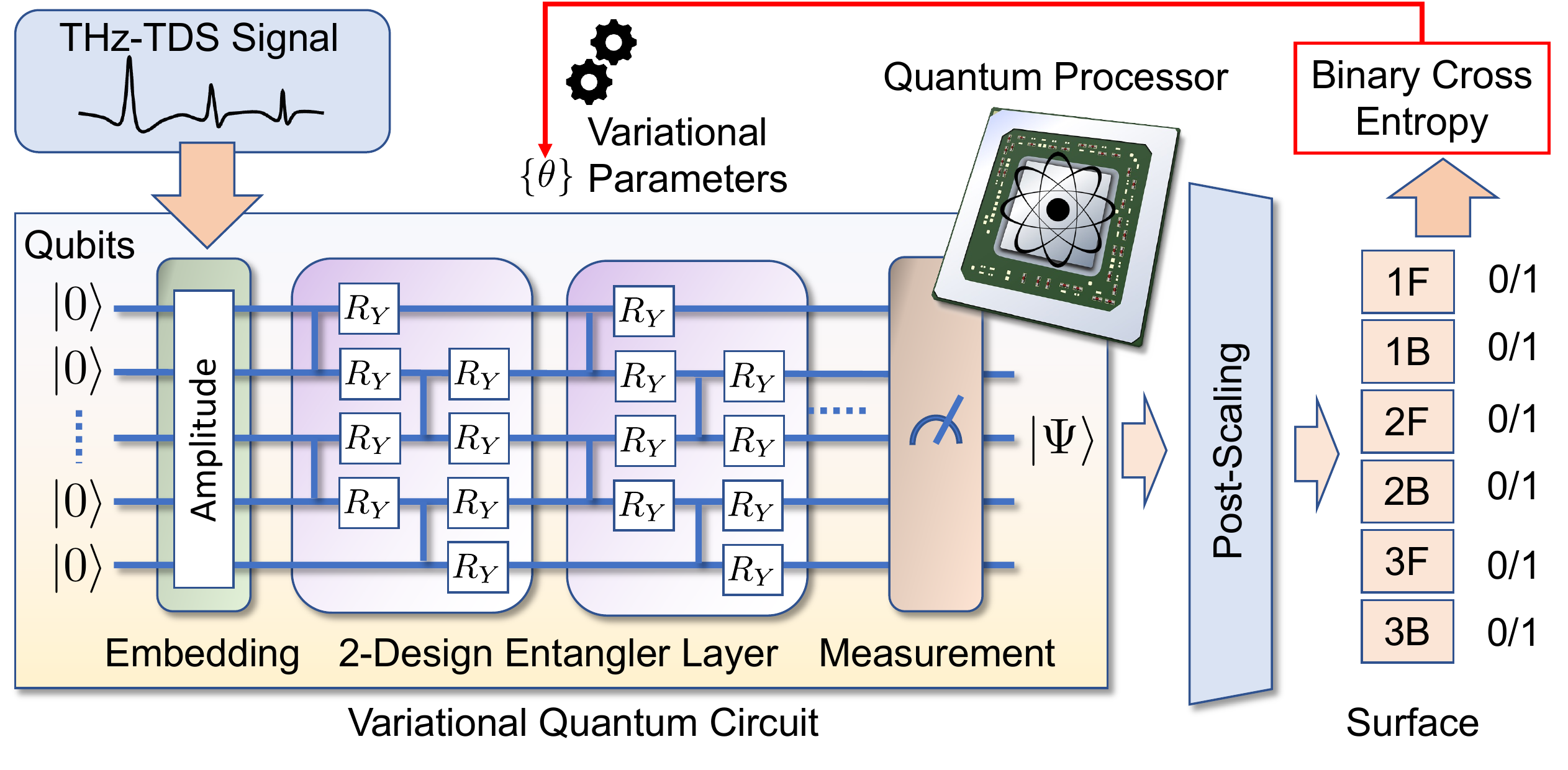}
		\caption{VQC feature extraction for THz waveform analysis.
		VQC employs amplitude embedding and 2-design ansatz to convert THz-TDS waveform into a feature vector.
		For 3-layer sample, $6$ binary classification scores for all front and back surfaces are computed for the model output.}
		\label{fig:qnn}
	\end{figure}

We consider a variational quantum circuit (VQC) to extract features as shown in Fig.~\ref{fig:qnn}. 
The VQC uses amplitude embedding to prepare quantum state according to THz-TDS waveform-domain signal, and the state evolves through a 2-design ansatz~\cite{cerezo2021cost}, which can approximately represent arbitrary unitary rotations, whose size is exponential to the number of variational angle parameters.
The advantage of the VQC lies in the fact that even small number of qubits and trainable parameters can support a long waveform (e.g., $10$-qubit circuits can encode $1024$-sample signals).
After quantum measurement, the extracted features are post-scaled or processed by additional classifier.
We then try to optimize VQC by minimizing the binary cross-entropy loss on each surface of material layers to address the shadow effect and contents recovery.

The QML feeds THz-TDS waveform-domain signal with a size of $196$-sample points, which are embedded into quantum amplitudes of $8$-qubits QPU. 
The quantum states are modified by the 2-design ansatz having $2$ layers of staggerred Pauli-Y rotations and controlled-Z entanglers. 
Finally, the quantum state is measured as a transformed feature vector of size $196$.
We further use a deep neural network (DNN) model to analyze the feature vector for the classification.
The DNN model uses $5$ linear layers with Mish activations and batch normalizations, whose node sizes are halved successively.
The number of trainable parameters is $28$ and $36{,}674$, respectively for QML and DNN models.
They are trained over $1000$ epochs by the stochastic gradient descent algorithm with a learning rate decaying by $0.5$ every $10$ epochs from the initial value of $5.0$.
The minibatch size is chosen to be $128$.

	\begin{figure}[t]
		\centering
		\includegraphics[width=0.9\linewidth]{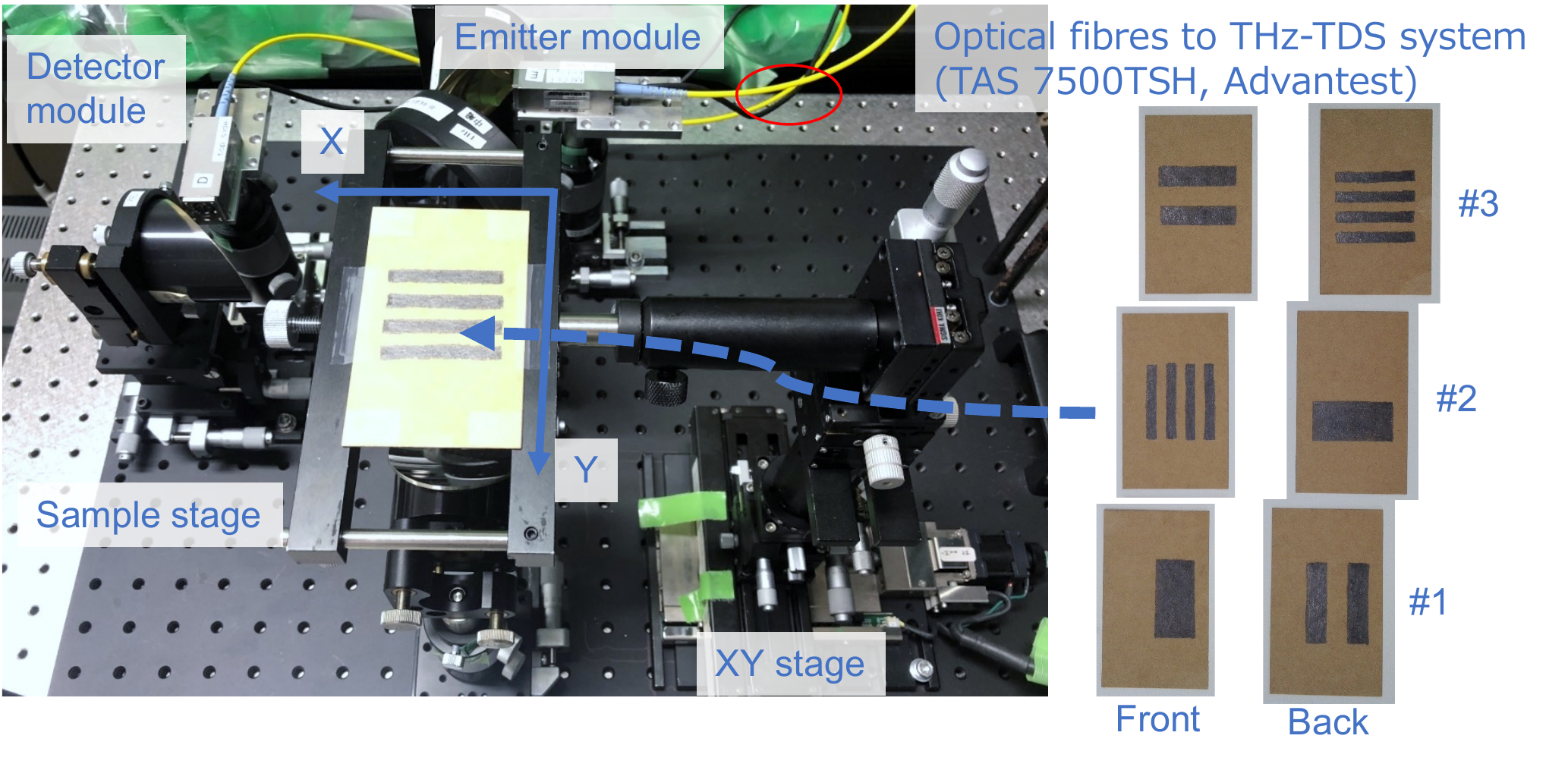}
		\caption{THz-TDS multi-layer imaging testbed for raster scan of
		double-sided papers.}
		\label{fig:system}
	\end{figure}

	\begin{figure}[t]
		\centering
		\subfloat[][DNN]{
			\includegraphics[width=0.46\linewidth]{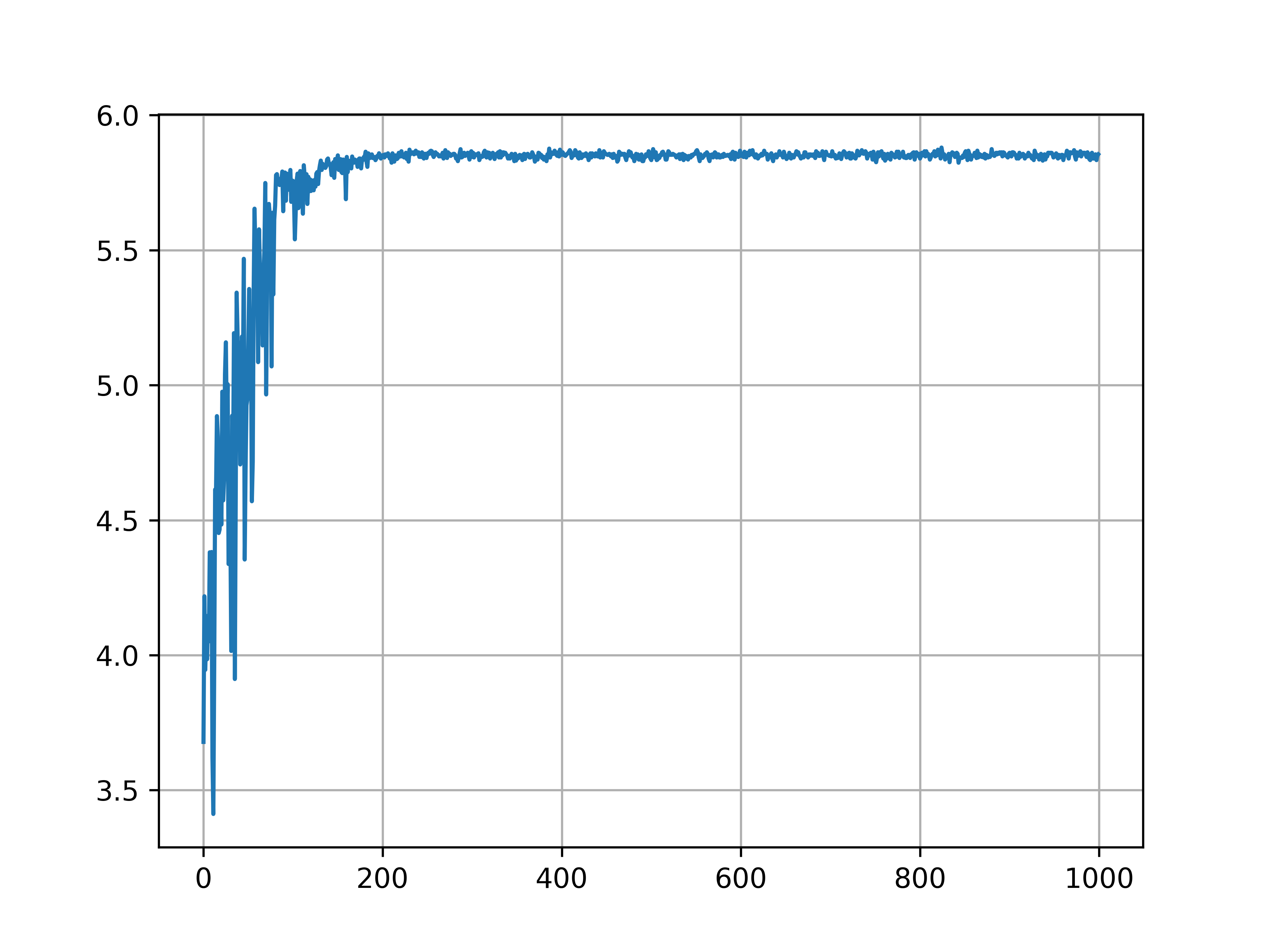}
		}
		\hfil
		\subfloat[][QML+DNN]{
			\includegraphics[width=0.46\linewidth]{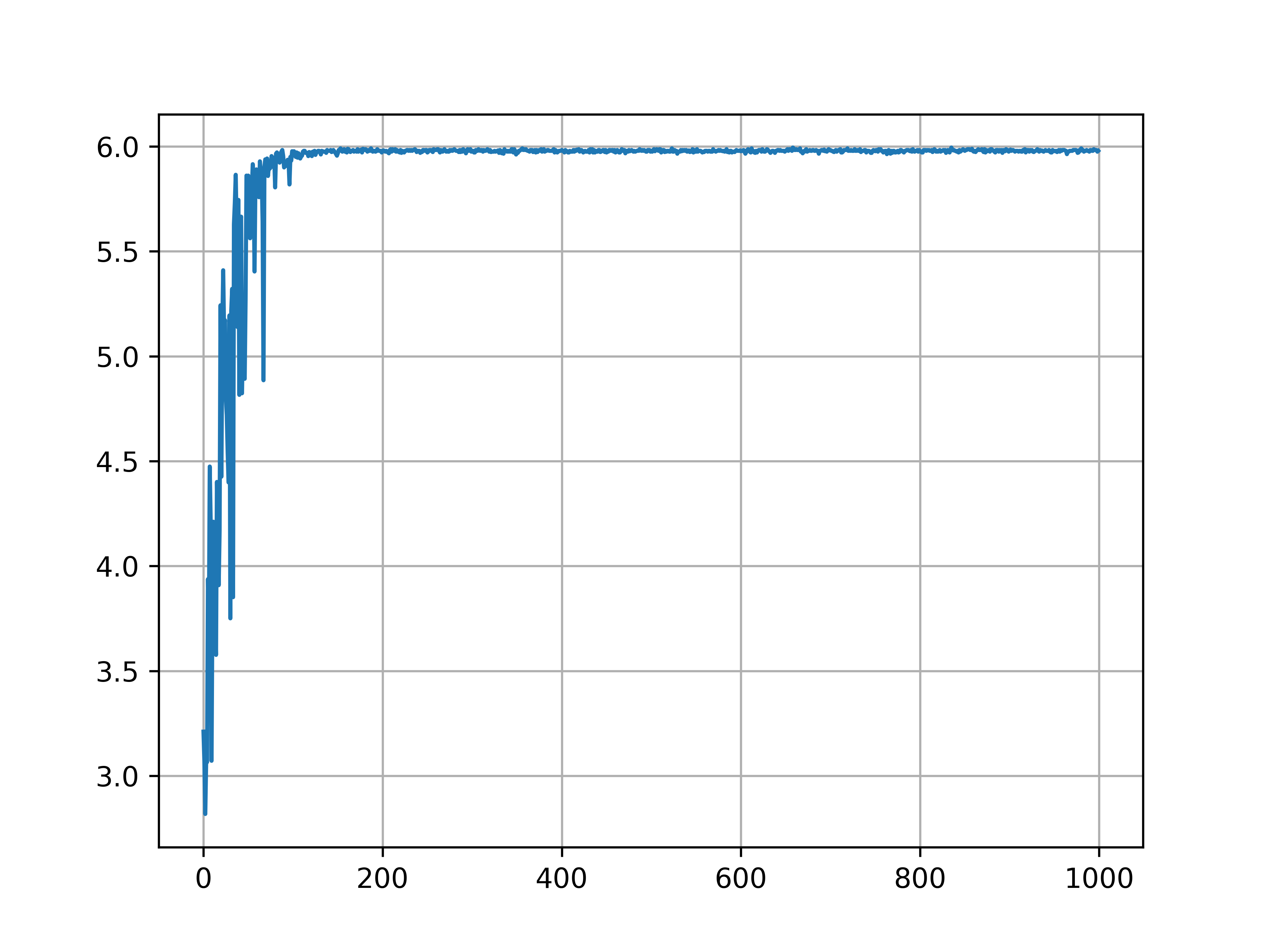} 
		}
	\caption{Training trajectory of classification accuracy over epochs for validation dataset.}
	\label{fig:epoch}	 
	\end{figure}

	\begin{figure}[t]
		\centering
		\includegraphics[width=\linewidth]{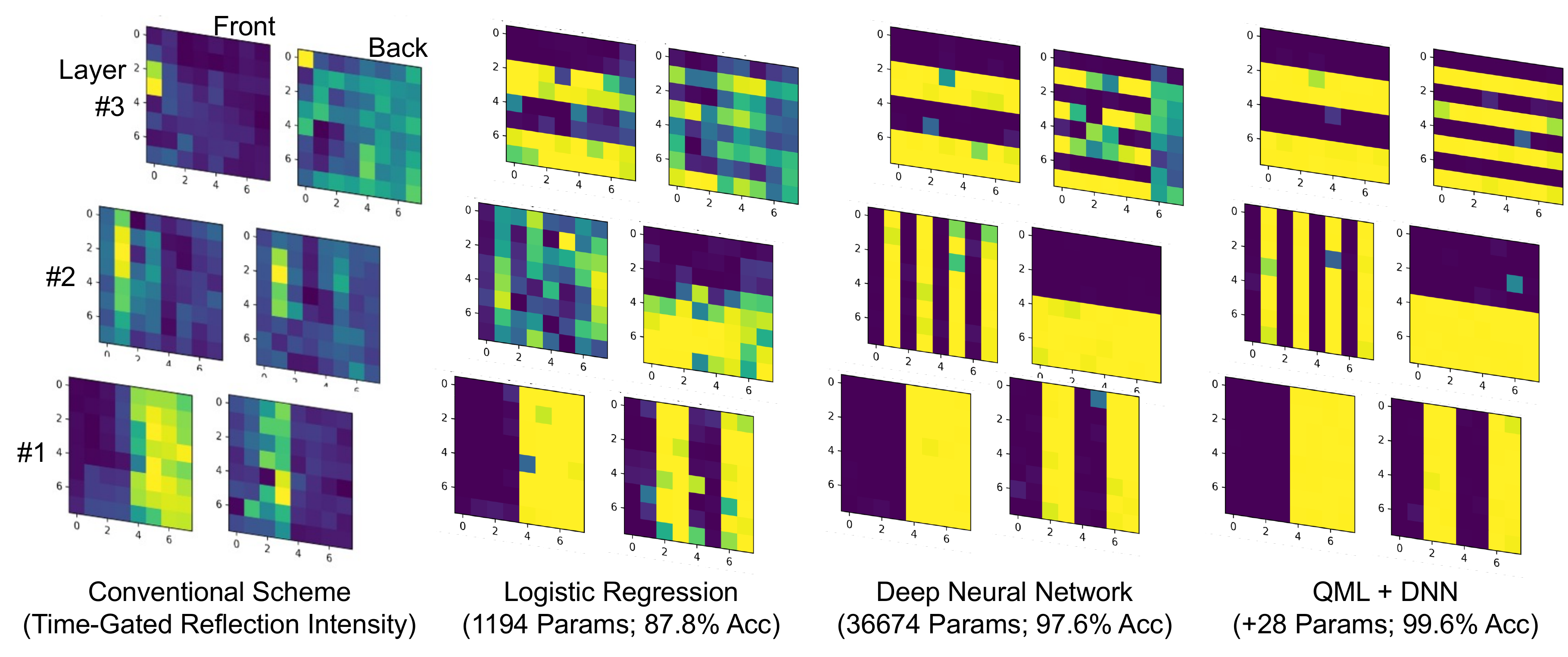} %
		\caption{THz imaging recovery performance comparison: 
		reflection intensity method; 
		logistic regression method;
		DNN method;
		QML+DNN method.}
		\label{fig:results}		 
	\end{figure}
	
\section{Experimental Validation}

	Fig.~\ref{fig:system} shows our THz-TDS testbed, where a $3$-layer sample is mounted on the raster scanning stage. 
	Both front and back surfaces of each layer are drawn with pencils to cover an area of $40 \times 40$~mm$^2$ that is divided into $8 \times 8=64$ pixels. 
	Each pixel of the size $5 \times 5$~mm$^2$ is associated with a unique binary label $\bf{c}$ according to the pencil drawings. 
	For instance, ${\bf c}=[1, 0, 1, 0, 1, 0]$ implies that all front surfaces are covered by the drawing while the back surfaces are blank. 
	With a scanning stepsize of $0.5$~mm, we have a set of $10 \times 10 = 100$ THz-TDS waveforms for each pixel and then randomly split them into the training ($60$\%), validation ($10$\%), and test ($30$\%) datasets. 

	The learning trajectory in terms of the total classification accuracy over $6$ surfaces is shown in Fig.~\ref{fig:epoch} over epochs. 
	It is seen that the QML-based model converges at around 100 epochs, which is faster than the DNN-based model.
	Moreover, the QML method achieves nearly perfect prediction accuracy.

	Fig.~\ref{fig:results} shows the comparison of the traditional time-gated reflection intensity approach with several learning-based approaches. 
	The traditional approach suffers from the shadowing effect from the front layers to the deep layers and the limited separation between two closely spaced surfaces. 
	On the contrary, the learning-base methods show better content recovery over the $6$ surfaces with a reduced shadowing effect. 
	Nevertheless, deeper layers show less confident scores than the front layers as expected. 
	The state-of-the-art DNN method having $36{,}674$ trainable parameters significantly outperforms the classical logistic regression method.
	Whereas, the proposed QML-based method having just $28$ VQC parameters achieves an excellent accuracy of $99.6\%$ to assist the DNN classifier.

\section{Conclusions}

This paper proposed the first proof-of-concept QML-based THz multi-layer content extraction for high-capacity 3D positioning. 
The proposed quantum feature extraction method integrated with DNN model was experimentally verified to improve performance against depth variations and shadowing effect for double-sided $3$-laminate imaging.

\bibliographystyle{IEEEbib}
\bibliography{bibs/bib-THz-2020}

\end{document}